\documentclass[journal]{IEEEtran}
\usepackage{amsmath,amsfonts}
\usepackage{algorithmic}
\usepackage{algorithm}
\usepackage{array}
\usepackage[caption=false,font=normalsize,labelfont=sf,textfont=sf]{subfig}
\usepackage{textcomp}
\usepackage{stfloats}
\usepackage{url}
\usepackage{verbatim}
\usepackage{graphicx}
\usepackage{cite}
\usepackage{orcidlink}
\usepackage{hyperref}
\usepackage{textgreek}
\hypersetup{
    colorlinks=true,
    linkcolor=blue,
    filecolor=magenta,      
    urlcolor=blue,
    citecolor=blue
    }
\begin{document}

\title{Updated electrical design of the Diagnostic Neutral Beam Injector in RFX-mod2}

\author{Marco Barbisan\orcidlink{0000-0002-4065-3861},~\IEEEmembership{member,~IEEE}, Bruno Laterza\orcidlink{0000-0002-4767-979X}, Luca Cinnirella\orcidlink{0009-0009-3857-7159}, Lionello Marrelli\orcidlink{0000-0001-5370-080X}, Federico Molon\orcidlink{0000-0001-7117-6431}, Simone Peruzzo\orcidlink{0000-0003-3626-1707}, Enrico Zampiva\orcidlink{0000-0002-7516-3731}

\thanks{This paper was produced by the IEEE Publication Technology Group. They are in Piscataway, NJ.}
\thanks{Manuscript received XXXXXX, 2025; revised XXXXXX, 2026.}
\thanks{M. Barbisan, L. Marrelli, S. Peruzzo and E. Zampiva are with CNR, Istituto per la Scienza e la Tecnologia dei Plasmi (ISTP), C.so Stati Uniti 4, 35127, Padova, Italy (email: marco.barbisan@istp.cnr.it). }
\thanks{B. Laterza, L. Cinnirella, L. Marrelli, F. Molon and E. Zampiva are with Consorzio RFX (CNR, ENEA, INFN, Università di Padova, Acciaierie Venete SpA), C.so Stati Uniti 4, 35127, Padova, Italy.}
\thanks{L. Cinnirella is with Università degli Studi di Padova, Centro Ricerche Fusione (CRF), C.so Stati Uniti 4, 35127, Padova, Italy.}
}

\markboth{IEEE Transactions on Plasma Science,~Vol.~XX, No.~X, XXX~2026.}%
{Shell \MakeLowercase{\textit{et al.}}: A Sample Article Using IEEEtran.cls for IEEE Journals}

\IEEEpubid{0000--0000/00\$00.00~\copyright~2026 IEEE}

\maketitle

\begin{abstract}
The Diagnostic Neutral Beam Injector (DNBI) of the RFX-mod2 experiment (Consorzio RFX, Padova) is expected to provide novel and significant information about the Reversed Field Pinch confinement of fusion plasmas. The present DNBI, built by the Budker Institute of Plasma Physics, features an arc discharge H\textsuperscript{+} source, coupled to a 4-grid 50 keV acceleration system, to produce a 50 ms, 5 A ion beam.
This contribution presents the latest upgrades of the DNBI. The High Voltage Deck (HVD) was completely restructured, and the power transfer was simplified to a single phase insulation transformer. The 50 kV distribution circuit was modernized and made safer against breakdowns.  Several custom power supplies in the HVD were designed and procured; their electronic boards were developed to be multipurpose in the DNBI, simplifying the system and improving its maintainability. The features of the magnetic insulation power supply and gas valve power supplies are presented in detail. Finally, the new PLC control system was improved for better protection of the CPU against overvoltages and for better scalability and maintainability of the system. 
\end{abstract}


\IEEEpubidadjcol

\section{Introduction}
\IEEEPARstart{T}{he} RFX-mod2 experiment, now in the procurement and assembly phase, is the upgrade of the previous RFX-mod machine, in Consorzio RFX, Padua.  While being a flexible tool to explore magnetic confinement configurations for fusion, RFX-mod2 will be mainly devoted to study Reversed Field Pinch (RFP) plasmas, thanks to its improved passive stabilization properties at the plasma edge \cite{Marrelli2019, Peruzzo2023, Terranova2024}. Among the several diagnostics, Charge eXchange Recombination Spectroscopy (CXRS) and Motional Stark Effect (MSE) will allow to measure ion temperature, ion flow, magnetic field intensity and magnetic field direction. The main target of these diagnostics is the plasma core, where edge diagnostics cannot provide reliable measurements; while having been extensively applied to Tokamaks, less experience is available with RFPs \cite{Carraro1999, Kuldkepp2006, Ko2010, DenHartog2011}. 

The CXRS and MSE measurements rely on emissions that result from the active interaction between the plasma and a neutral beam of H or D atoms. At Consorzio RFX, RFX-mod was equipped in 2005 with a Diagnostic Neutral Beam Injector (DNBI),  provided by the Budker Institute for Nuclear Physics (BINP, Novosibirk) \cite{Korepanov2004}. It was based on an arc H\textsuperscript{+}/D\textsuperscript{+} ion source. Ions were extracted and accelerated to 50 kV, for an equivalent current of 5A, by a system of four grids: Plasma Grid (PG), Extraction Grid (EG), 3\textsuperscript{rd} grid (a repeller for backwards-traveling electrons) and Grounded Grid (GG). Ions were neutralized by charge exchange with gas molecules; the resulting neutral beam had a maximum equivalent current of 2 A. Residual ions were deflected and dumped by the magnetic field of a Helmholtz coil. The neutral beam could last 50 ms per pulse, with the possibility of modulating the beam up to 250 Hz to separate the active emissions from the passive ones, that are expected to be non negligible \cite{Carraro1999}. 

The DNBI power supply system and the control and data acquisition system are now obsolete and not in the conditions to operate reliably, especially considering the 50 kV potential required to accelerate the ions. From 2023, a work of redesign, simplification and upgrade of the electrical systems has been carried out. Any collaboration with BINP was not possible due to geo-political issues. The new design of the DNBI-related electric devices was presented in ref. \cite{DNBI2025}.

From that moment, the redesign activities continued: the design of the High Voltage Deck (HVD) was finalized, as well as the detailed design of several devices and their electronic boards; a review of the control and data acquisition system led to changes in its general architecture. This paper presents the outcome of these design activities, in the hope to provide useful information for other facilities hosting similar BINP DNBIs \cite{Wood2023, Listopad2024, Mysiura2025}. The practical implementation of the design is shown, too.
\IEEEpubidadjcol

\section{High Voltage Deck Improvements}
\subsection{Overview}
The High Voltage Deck (HVD) is the DNBI structure that stays at the same potential of the source during beam extraction and acceleration, i.e. at +50 kV. It contains all the devices that serve the ion source and the Plasma Grid (PG, the grid facing the source), plus the High Voltage (HV) components to derive the voltage for the Extraction Grid (EG, the grid following the PG) and overvoltage protection components. Due to its obsolescence, to HV holding issues and to the faulty oil insulation transformer powering it, the old HVD was prioritized in the process of upgrading the DNBI. 

\begin{figure}[htbp]
\centering
\includegraphics[width=8.5cm]{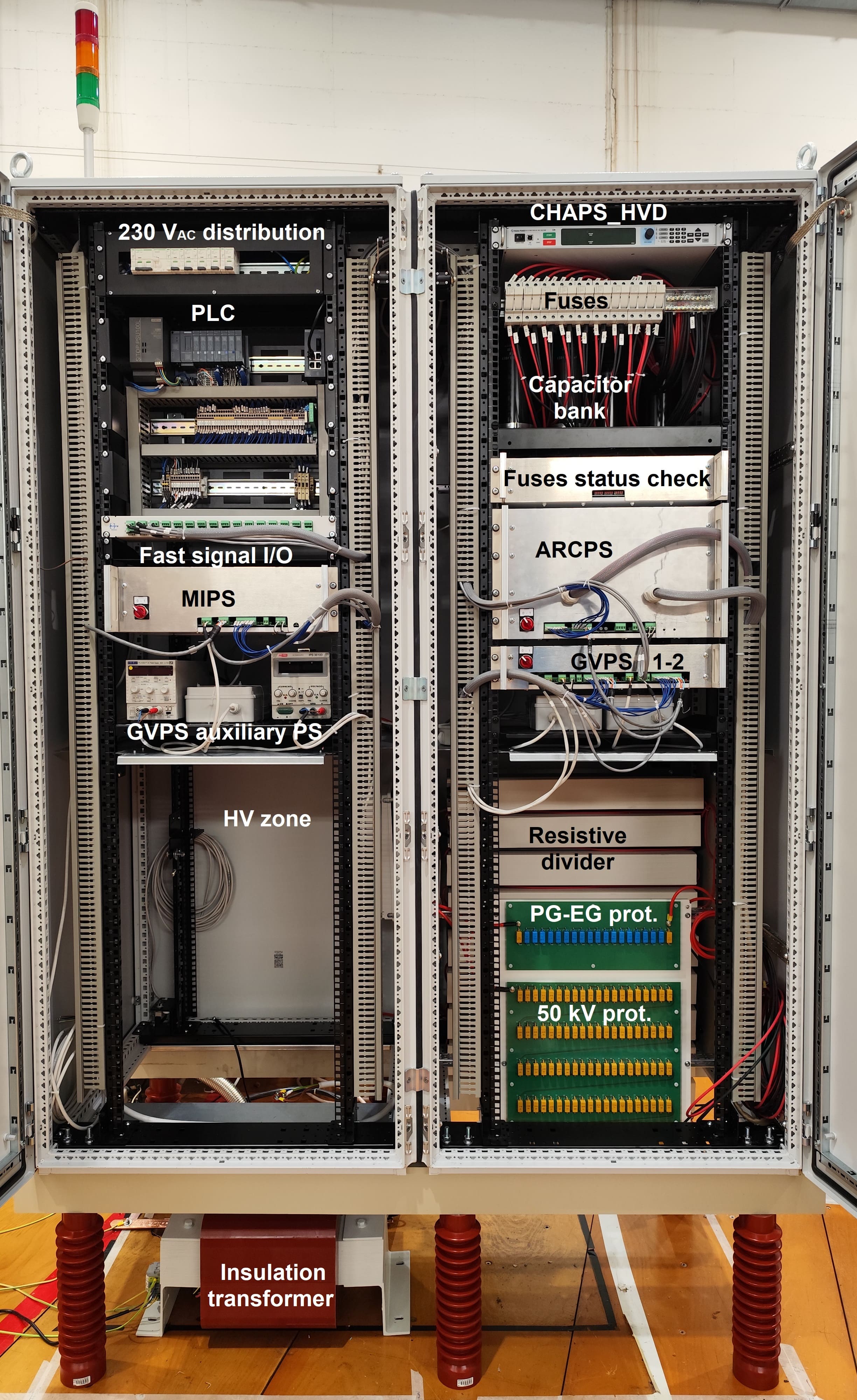}
\caption{Implementation of the HVD design (final cabling and setup of HV circuit is still ongoing).}
\label{fig:HVDreal}
\end{figure}

While the old HVD, sitting on top of the insulation transformer, was enclosed in grounded cabinets together with HV components, the new HVD consists of two cabinets, raised 46 cm from the floor (the wooden mezzanine around the experiment) with HV insulators and PP-H bars. The recent implementation is shown in fig. \ref{fig:HVDreal}. The protection of nearby instruments will be ensured by a plastic cage (to be installed in future); human safety will be guaranteed by a disconnector switch, grounding the HVD as long as the experimental hall is accessible to personnel. This choice allows much more space in the HVD, and the possibility of organizing devices in standard rack and cabinets, which is more rational and maintainable. The custom triple insulation transformer in oil was replaced with a standard resin single phase insulation transformer (230 V\textsubscript{AC}, 5 kVA, primary-secondary insulation tested at 70 kV RMS for 1 min.), simplifying the system and improving its maintainability. 

Following the original upgrade concept in ref. \cite{DNBI2025}, and as shown in fig. \ref{fig:HVDreal}, the HVD is internally organized in a bottom section, for the HV components, and in a top section, hosting the controls and source-related devices:
\begin{itemize}
    \item 230 V\textsubscript{AC} distribution system.
    \item PLC and fast data I/O (sec. \ref{sec:DAQ}).
    \item Magnetic Insulation Power Supply (MIPS, sec. \ref{sec:MIPS}), to power a solenoid which should generate an axial magnetic field in the ion source, reducing plasma losses.
    \item CHAPS\_HVD (0-1000 V\textsubscript{DC}, 0-1.5 A), to load the 54 mF, 900 V capacitor bank (6 sections, each protected by a fuse) for the Arc Power Supply (ARCPS). 
    \item A protection board, to check the status of the fuses for each one of the 6 sections of the capacitor bank, together with a safety relay, discharging the capacitor bank on a 120 \textOmega, 50 W resistor (6.5 s 1/e discharge time) in case of emergency.
    \item ARCPS, designed to ignite the arc in the ion source with max. 2 kV, and sustain the discharge at max. 150 V, 500 A \cite{DNBI2025}.
    \item Gas Valve Power Supplies (GVPS) 1-2 (sec. \ref{sec:GVPS}), to independently feed hydrogen into the source from the cathode and the anode.
\end{itemize}
The HVD and the insulation transformer were successfully tested to stand 100 kV for 1 min..

\subsection{High Voltage Components}
\begin{figure*}[htbp]
\centering
\includegraphics[width=15cm]{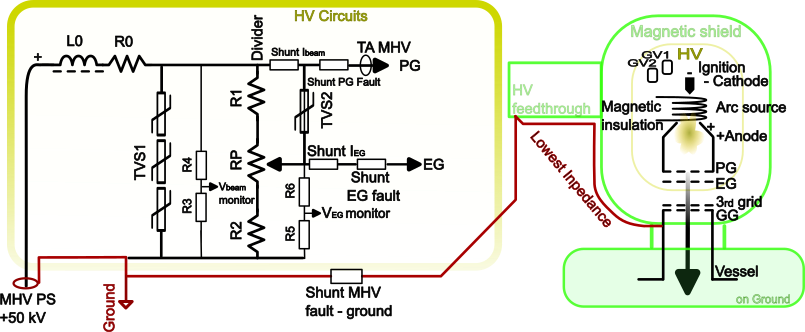}
\caption{High voltage circuit in the HVD.}
\label{fig:HVDfinal}
\end{figure*}

\begin{figure}[htbp]
\centering
\includegraphics[width=8.5cm]{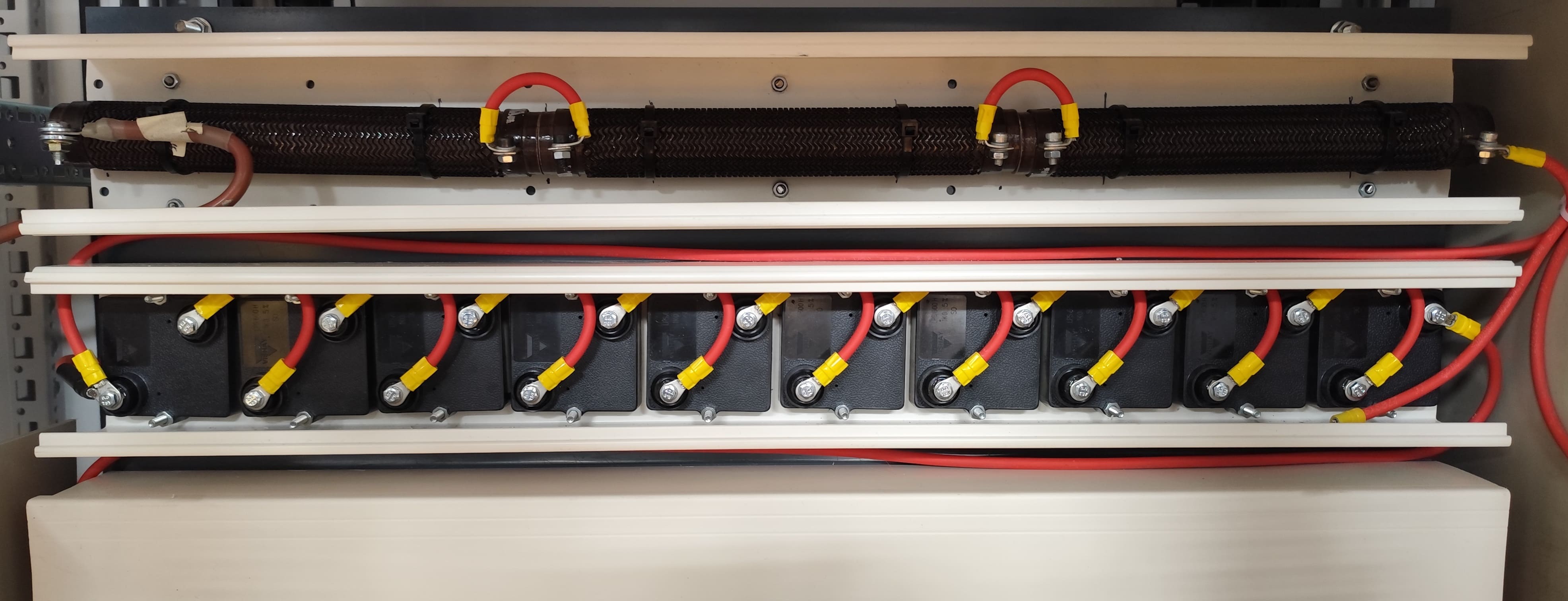}
\caption{R0 and first section of the resistive divider (R1).}
\label{fig:divider}
\end{figure}

The HV components in the HVD are schematized, together with the ion source and the acceleration system, in fig. \ref{fig:HVDfinal}. The output of the High Voltage Modulator (MHV) reaches the HVD chassis through a coaxial cable. The MHV output V\textsubscript{beam}=50 kV is protected from the effects of breakdowns by an inductor L0 and R0=300 \textOmega, composed of three wirewound (for extra inductivity) resistors RBEF0300100R0KFB00 (100 \textOmega, 300 W). The resistors are enclosed in a plastic case for further protection (fig. \ref{fig:divider}, top segment). 

The EG voltage V\textsubscript{EG} is derived from the MHV by means of a resistive divider (R1, R2, RP). It is composed of 80 Vishay LPS 300 resistors (thin film, 300 W max.), organized in 8 rows, each one enclosed in a plastic case for better cross-row insulation (e.g. fig. \ref{fig:divider}, bottom segment). All resistors are rated 1 k\textOmega, except in the second row, in which 100 \textOmega{} resistors are present. In this way R1=10 k\textOmega{} (row 1), RP=1 k\textOmega{} (row 2), R2=60 k\textOmega{} (rows 3-8). V\textsubscript{EG} will be derived by one of the resistor terminals in the second row, allowing a minimum V\textsubscript{EG} regulation step of 70 V. Resistors can be freely rearranged to get the best match for beam optics. In case of arcs, each resistor row can withstand a potential difference 50 kV within the 50 ms pulse duration.

\begin{figure}[htbp]
\centering
\includegraphics[width=8.5cm]{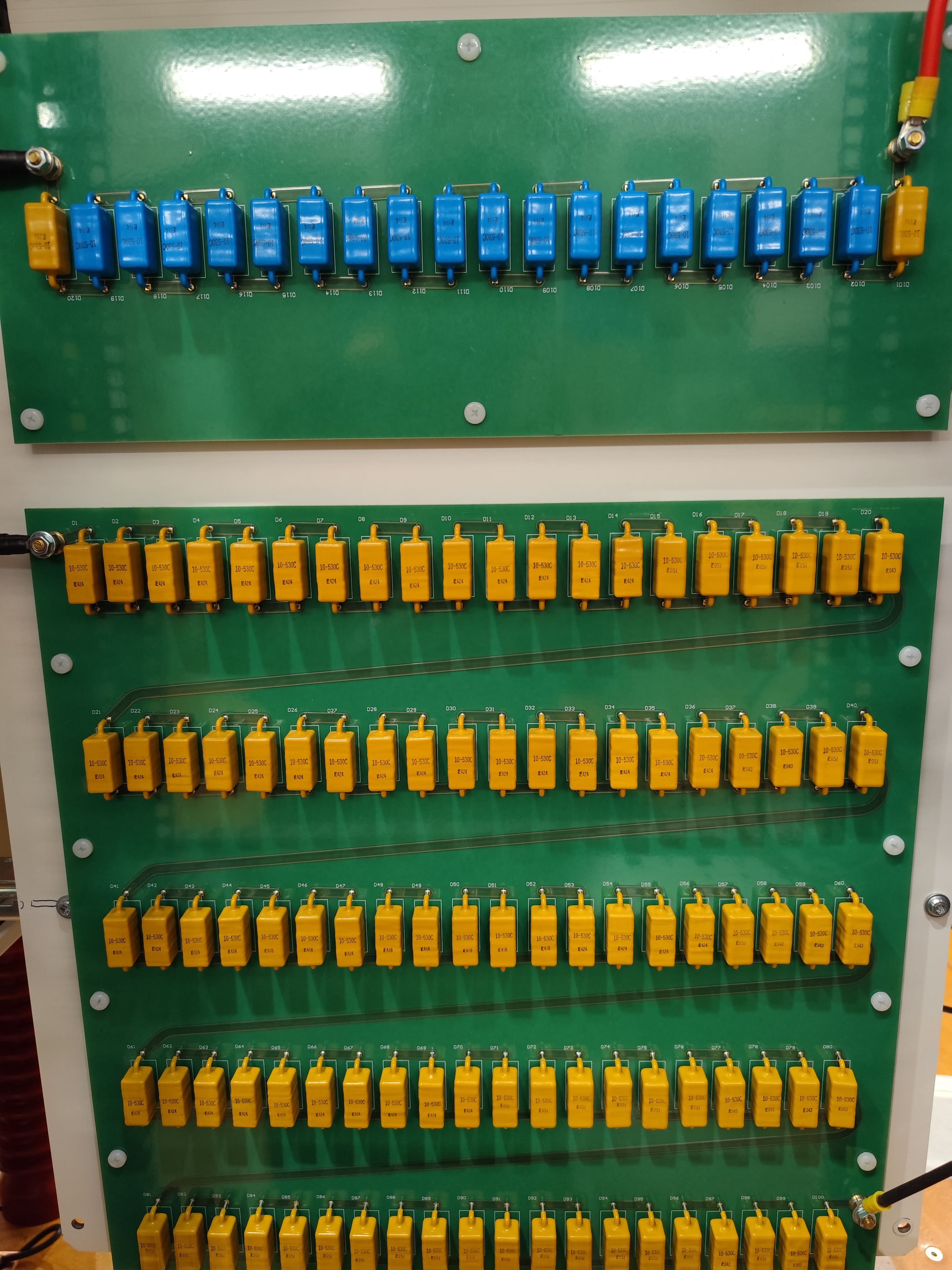}
\caption{TVS protection for 50 kV (TVS1, bottom) and for the PG-EG gap (TVS2, top).}
\label{fig:tvs}
\end{figure}

The voltage on the third grid, a repeller to block electrons traveling backward, will be provided by a dedicated -1 kV, 1 A power supply; it will not be placed in the HVD but in the ground cabinets.
TVS diodes are employed to provide overvoltage protection, both for the 50 kV output (TVS1) and for the single PG-EG gap (TVS2). Compared to the varistors used in the original HVD, TVS will offer a steeper response with the rise of voltage. The selected TVS, AK10-530C-Y, has a standoff voltage of 530 V, a reverse breakdown between 560 V and 619 V, and a peak rating of 750 V at 10000 A. In TVS2, 20 TVS in series are placed in a row to protect the PG-EG gap (fig \ref{fig:tvs} top PCB); the spacing between nearby electrodes is 13 mm (PCB traces). For V\textsubscript{beam}, in TVS1 100 TVS were arranged in 5 rows of 20 (fig. \ref{fig:tvs}, bottom PCB); the separation between electrodes of nearby rows is 47 mm (PCB traces). The output of each row was carried diagonally to the input of the following one, to limit the electric field and the risk of arcs. The effective number of TVS to be used will be fine-tuned during the first commissioning phase, to match the overall TVS response with the circuit safety needs. The TVS boards and the resistive divider boards, now fastened inside the HVD as shown in fig. \ref{fig:HVDreal}, will be lowered to the floor level to be grounded at the bottom. At last, the 3rd grid will be protected at ground voltage by a further set of TVS (not visible in fig. \ref{fig:HVDfinal} nor in the HVD) in case of breakdowns between the third grid and the Grounded Grid (GG), or more likely in case of breakdowns between the EG and the third grid. Diodes will also protect the third grid power supply against polarity reversal, in case of breakdowns.

Several auxiliary components will provide measurements and safety interlocks. Two resistive dividers will provide 0-10 V monitor signals for V\textsubscript{beam} and V\textsubscript{EG} in the 0-100 kV range; more specifically, R3=R5=2 k\textOmega{}, while R4=R6=20 M\textOmega{}. Shunts in the PG and EG terminals will measure the extracted beam current I\textsubscript{beam} and the beam current impinging on the EG in case of poor beam focusing, respectively. I\textsubscript{beam} will be also measured by an insulated Hall current sensor (LEM IT 150-S). Two other shunts, placed in series to the previously mentioned ones, will inform the fast section of the control system in case of breakdowns, triggering a temporary switch off of MHV and a temporary reduction of the arc current in the ion source. The two fault shunts will work together with a third one, which links the DNBI vessel, including the Grounded Grid (GG), to the common ground. All the fault shunts will be dimensioned for higher currents compared to the measurement shunts. The system of three fault shunts will be able to identify the locations of breakdowns: if between PG and EG, the respective fault shunts will be triggered. If a breakdown will happen in the PG-3rd grid gap (and in cascade to the GG), the EG fault shunt and the MHV fault shunt will be triggered.

\section{Upgrade of power supplies}
\subsection{Magnetic Insulation Power Supply}\label{sec:MIPS}
The MIPS, required to generate an axial magnetic field that reduces plasma losses, is hosted within the HVD, too. Its maximum ratings are 35 V and 15 A; while this can be accomplished by off-the-shelf commercial products, a dedicated design was carried out, since the MIPS should be able to generate arbitrary waveforms with high time resolution. For example, the MIPS output could be ramped during a 100 ms plasma pulse to compensate for plasma density drifts, probably due to variations in the gas pressure.

\begin{figure}[htbp]
\centering
\includegraphics[width=8.5cm]{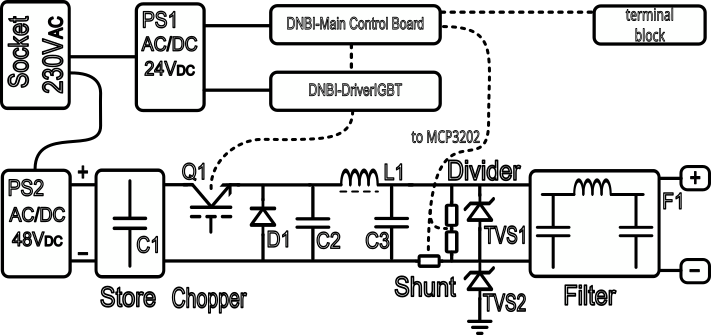}
\caption{Basic electric scheme of the MIPS.}
\label{fig:MIPSscheme}
\end{figure}

As schematized in fig. \ref{fig:MIPSscheme}, in the designed MIPS a 48 V\textsubscript{DC} power supply will load a 100 mF capacitor. The Q1 switch, a APT75GT120JU2 IGBT, will be controlled in pulse width modulation. L1=3mH, while C2=C3=1 \textmu F; the final filter is made by a 20 \textmu H and two 10 \textmu F capacitors. A resistive divider and a shunt provide voltage and current feedback for the control boards (sec. \ref{sec:boards}). Protection is provided by the diode D1 VS-U5FH120FA120, while TVS1 and TVS2 are 15KPA51A and 15KPA200A, respectively. At last, a 24 V\textsubscript{DC} power supply serves control boards (sec. \ref{sec:boards}).

\subsection{Gas Valve Power Supplies}\label{sec:GVPS}
\begin{figure}[htbp]
\centering
\includegraphics[width=8.5cm]{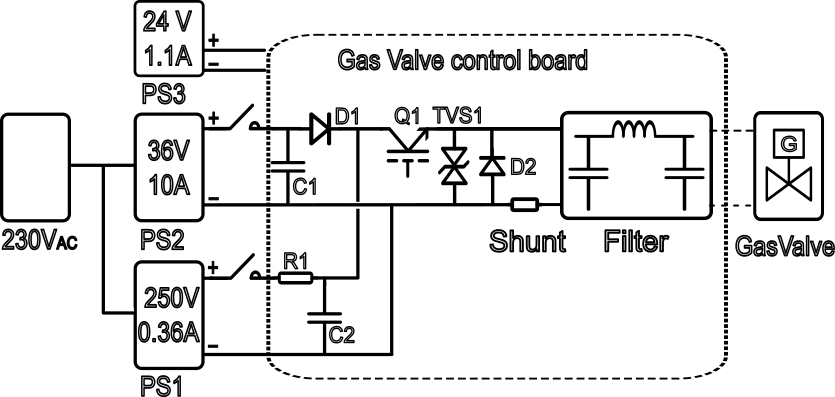}
\caption{Basic electric scheme of the GVPS.}
\label{fig:gvpsscheme}
\end{figure}

The GVPS power supplies control the solenoid valves at the anode and cathode of the ion source to regulate the gas inlet. They are "on-off" valves, only the timing and duration of the gas input can be set; the gas flux can instead be regulated by varying the upstream gas pressure. A mass flow controller is not placed in the ion source due to limited space and electromagnetic interference. 
Compared to the original power supplies, the new GVPSs have been made simpler and more flexible (fig. \ref{fig:gvpsscheme}). Besides the 24 V power supply for the control electronics (PS1), two auxiliary power supplies (visible in fig. \ref{fig:HVDreal}, mid height on the left) provide max. 250 V 0.36 A (PS1, Aim-TTi PLH250P) and max. 36V 10 A (PS2, RS-PRO IPS 3610D ). The first power supply provides the voltage for the valves to open quickly, the second one provides the sustain current to keep the valves open. The low voltage power supply is protected by a US5MC-HF diode. The switch is a STGW8M120DF3 IGBT, while the output filter is composed of a 20 \textmu H inductor and two 10 \textmu F capacitors. The IGBT is protected by the freewheeling diode D2 (same as D1) and the bidirectional TVS TVS1 (1.5KE300CA).

The time performance of the gas valve along with its power supply was measured on a test bench. With an opening voltage of 100 V and a sustain voltage of 24 V, a 4 ms delay between the rise of the control signal and the effective opening of the valve was measured, while a delay of 12 ms was present between the fall of the control signal and the effective closing of the valve. 

\subsection{Custom developed control boards}\label{sec:boards}
Besides the main power electronics, new control boards were designed to improve safety, reliability and controllability of the system. All of them use TRACO POWER TEN5-2423, TEN5-2421 and TEN5-2411, the only type of DC-DC converters which proved to stand RFX-mod2 magnetic field levels (130 mT in the test) \cite{DNBI2025}. Schematics, bill of materials, KiCad and gerber files are available in ref. \cite{schematics}.

\begin{figure}[htbp]
\centering
\includegraphics[width=8.5cm]{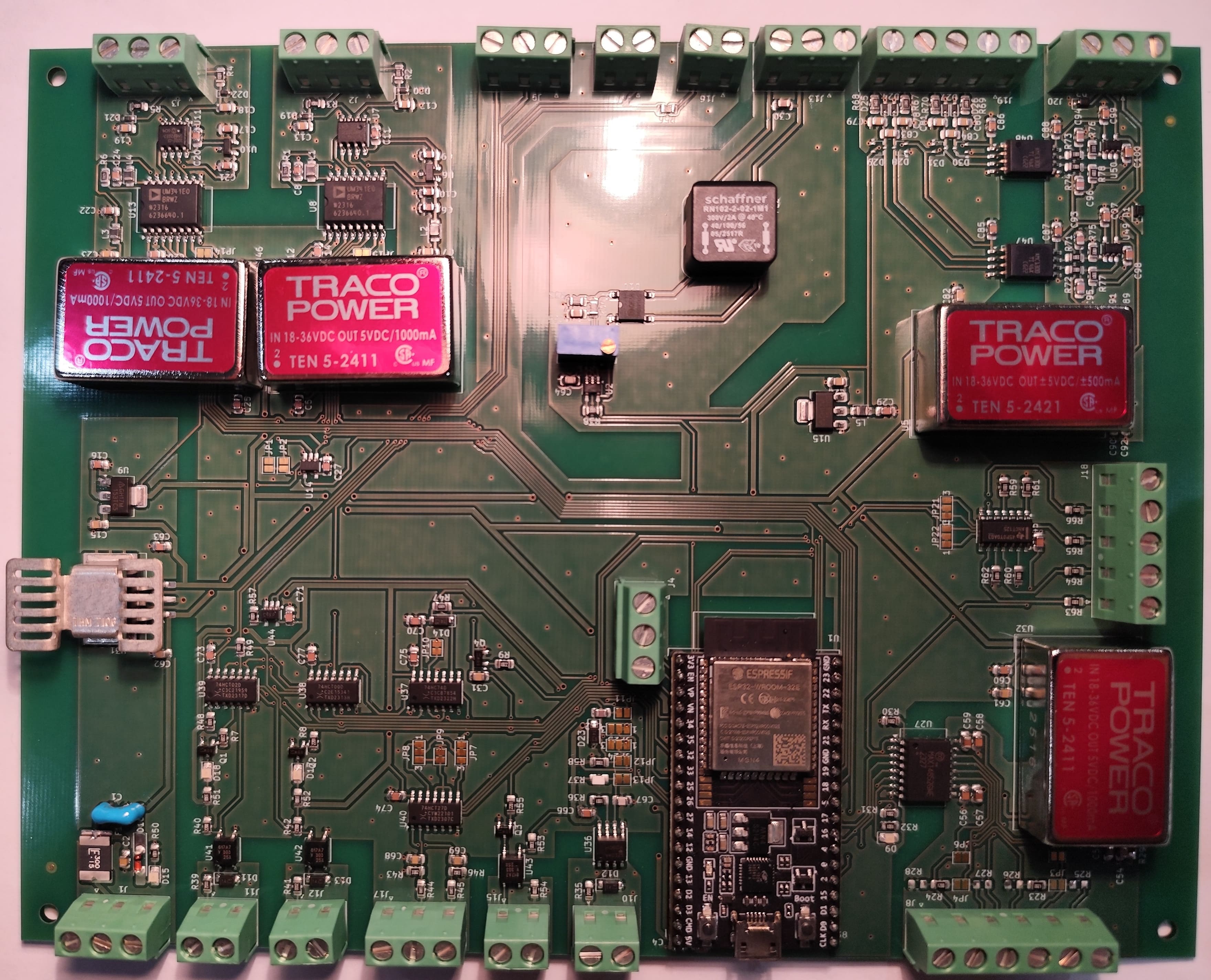}
\caption{The Main Control board.}
\label{fig:microesp32}
\end{figure}

The main control board (fig. \ref{fig:microesp32}) is based on the MicroESP32 micro-controller device to accomplish the following tasks:
\begin{itemize}
    \item Communicate with the PLC in the Profibus standard, to receive basic settings instructions and inform about the status of the device.
    \item Provide 20 kHz pulse width modulation signals for IGBTs.
    \item Receive voltage/current feedback through AMC1300DWV isolated amplifiers (5000 V\textsubscript{rms}) and process them through MCP3202 ADC converters (12 bit, 100 kS/s total).
    \item Manage IGBTs interlocks in the overall control chain. 
\end{itemize}

\begin{figure}[htbp]
\centering
\includegraphics[width=8.5cm]{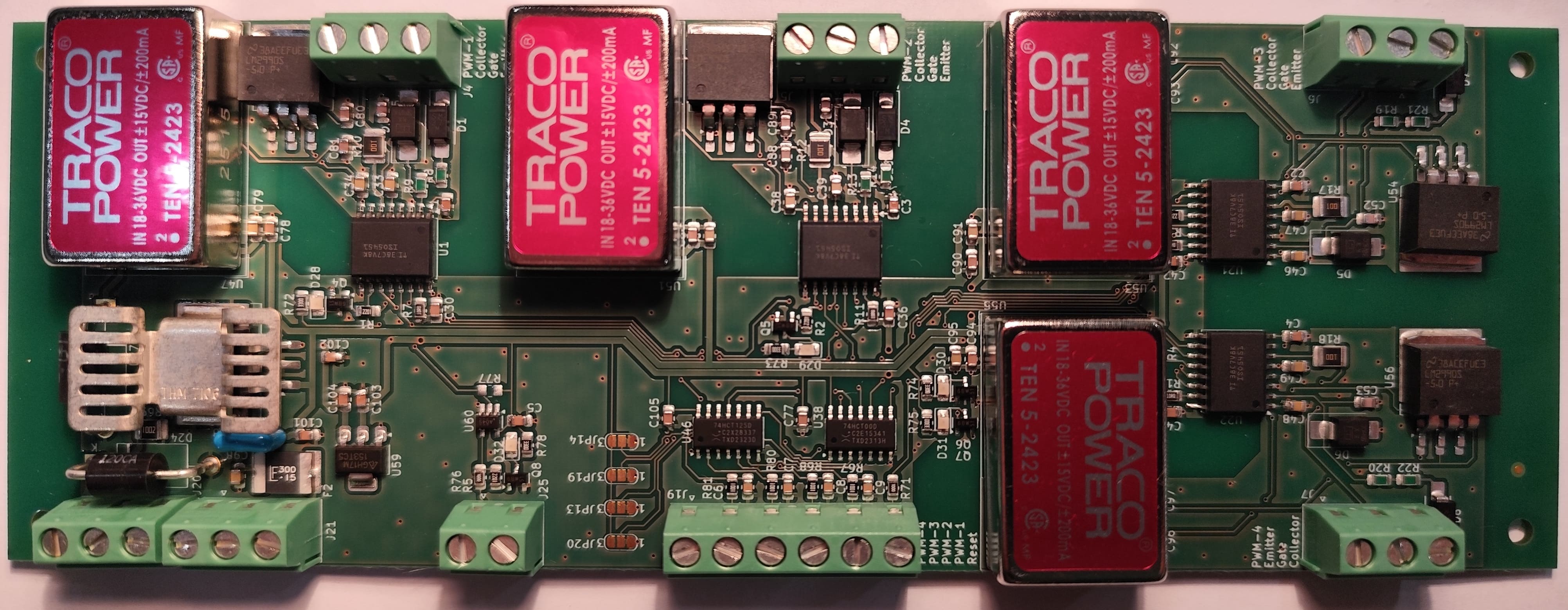}
\caption{The "DriverIGBT" board.}
\label{fig:driverigbt}
\end{figure}

The "DriverIGBT" board (fig. \ref{fig:driverigbt}) can control up to 4 IGBTs, using the ISO5451 gate drivers, which include desaturation and active Miller clamp protection system.

The Main Control board and the "DriverIGBT" board have been designed for accomplish multiple purposes in the DNBI. In the HVD, they will be part of MIPS and ARCPS. In the cabinets at ground potential, they will control the chains of voltage regulators and inverters that convert the DC voltage from the capacitor banks into AC for the MHV \cite{DNBI2025}. The possibility to use a unique set of boards for multiple DNBI devices improves the maintainability of the system.

\begin{figure}[htbp]
\centering
\includegraphics[width=8.5cm]{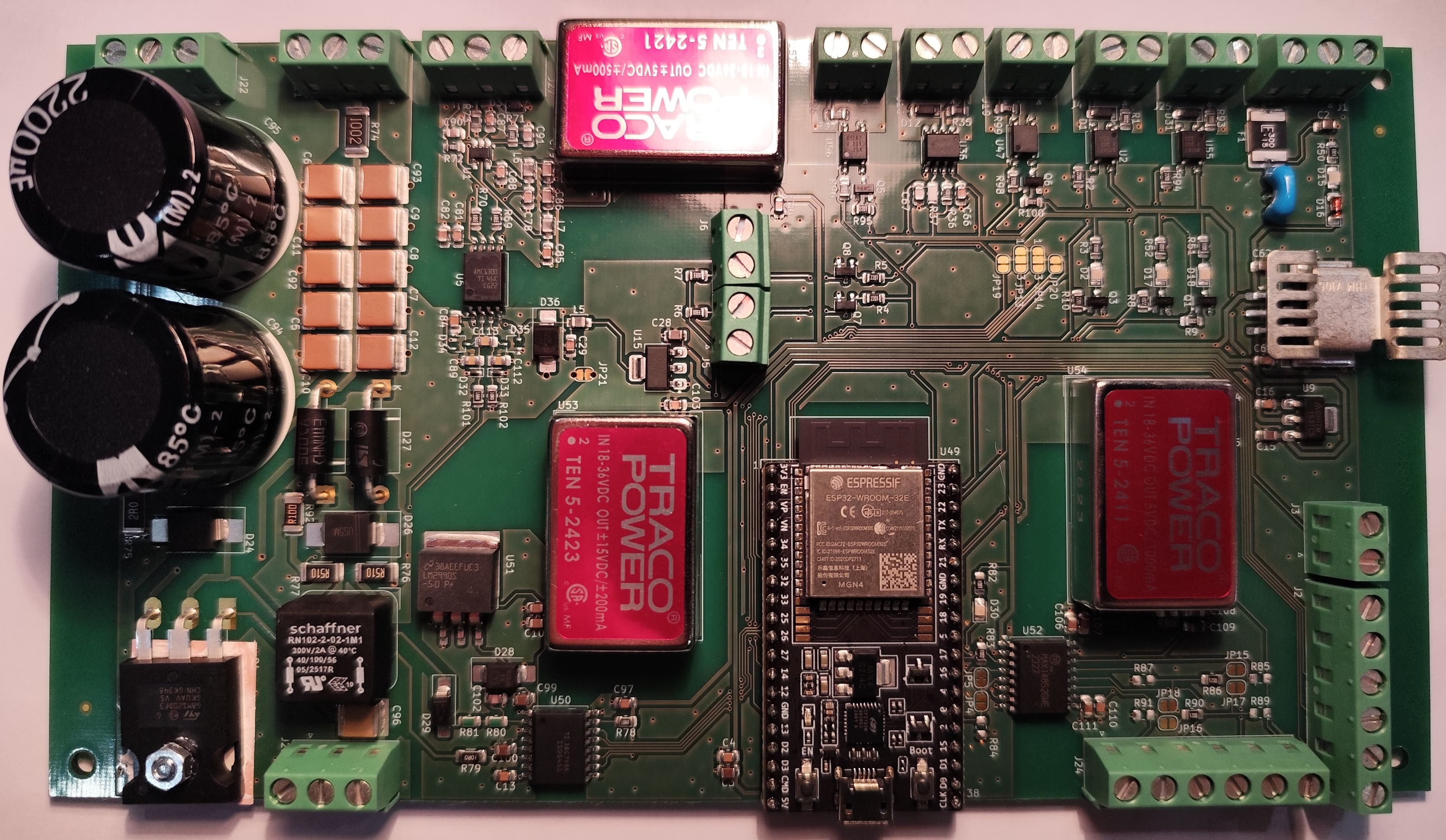}
\caption{The Gas Valve control board.}
\label{fig:gasvalve}
\end{figure}

At last, the Gas Valve control board (fig. \ref{fig:gasvalve}) implements all the circuitry of the device except the external power supplies. The board includes a MicroESP32 micro-controller for PLC communication, AMC1300DWV for safe acquisition of feedback signals, and the STGW8M120DF3 IGBT.


\section{Improvements on Control and Data Acquisition}\label{sec:DAQ}
\begin{figure}[htbp]
\centering
\includegraphics[width=8.5cm]{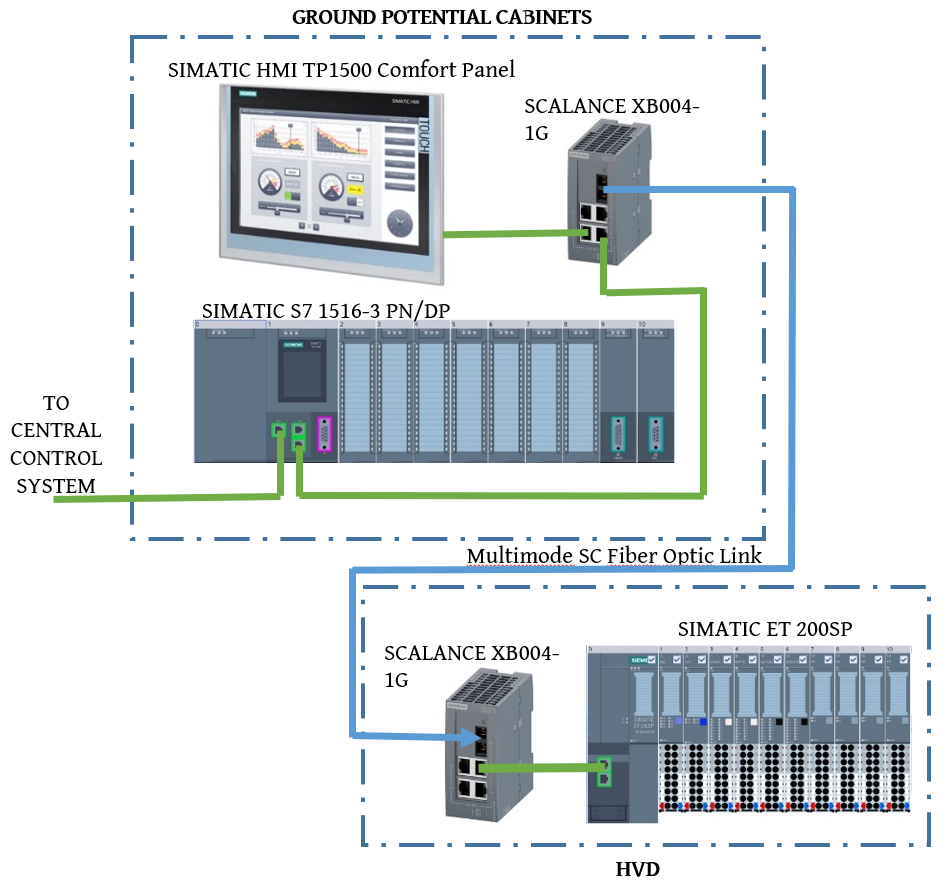}
\caption{Initial design for the DNBI PLC \cite{DNBI2025}.}
\label{fig:PLCold}
\end{figure}

As originally planned in ref. \cite{DNBI2025}, the DNBI CAMAC control system should be replaced by a modern PLC system to provide slow (below 50 S/s) analog and digital inputs-outputs; the PLC would implement all the safety measures for devices and personnel. The PLC would be complemented by a separate D-tAcq system for fast (1 MS/s) analog and digital inputs-outputs. The original plan for the PLC (fig. \ref{fig:PLCold}) foresaw a Siemens SIMATIC S7 1516-3 PN/DP CPU, directly interfaced to I/O boards for the low voltage cabinets. A SCALANCE XB004-1G switch would then connect the CPU to a SIMATIC HMI TP1500 panel for human interaction at ground potential, and to the HVD via fiber optic to provide high voltage insulation. Inside the HVD, another XB004-1G switch would transfer data to and from a SIMATIC ET200SP I/O interface.

\begin{figure}[htbp]
\centering
\includegraphics[width=8.5cm]{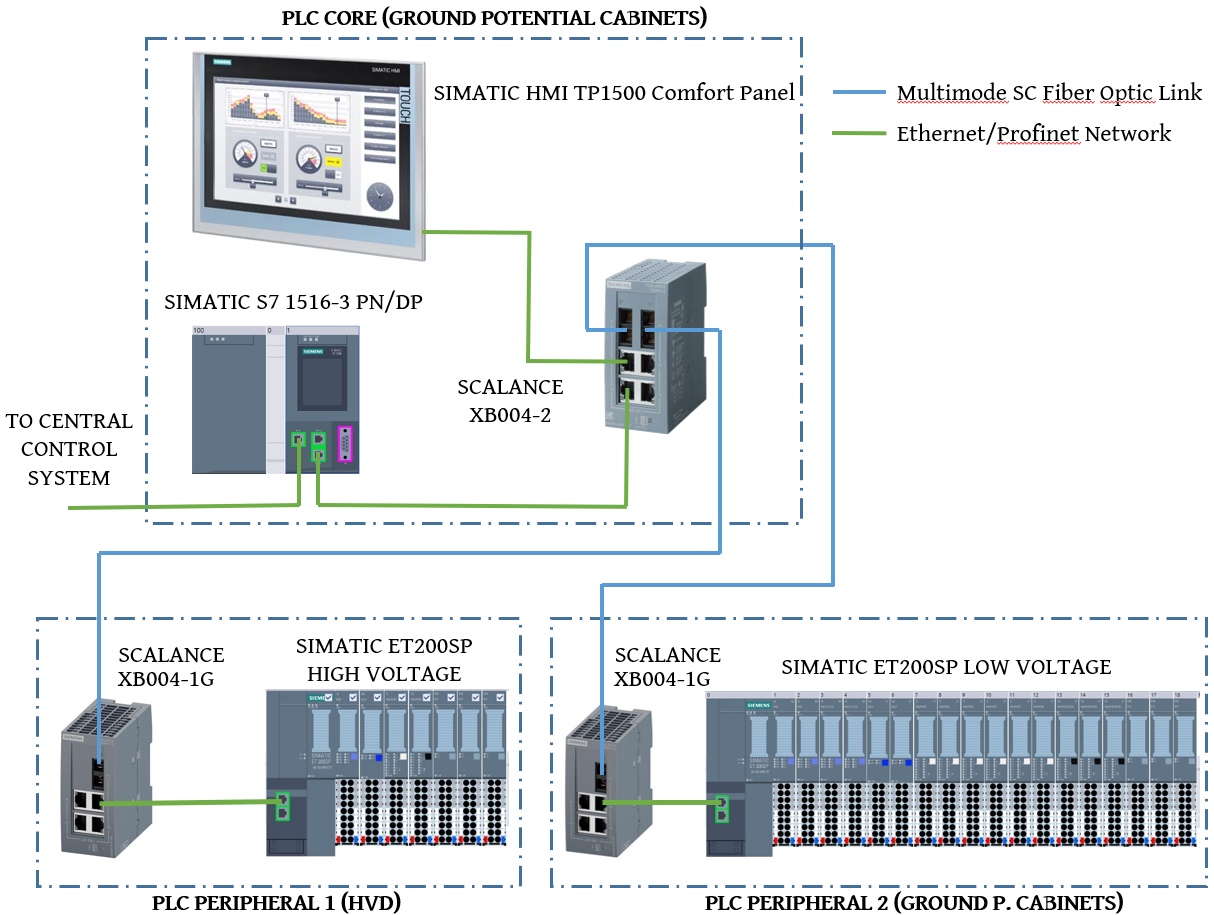}
\caption{New scheme for the DNBI PLC.}
\label{fig:CODACnew}
\end{figure}

In the new configuration (fig. \ref{fig:CODACnew}), the I/O at ground potential is handled by another ET200SP peripheral, connected to the CPU via fiber connection, similarly to what was implemented in the HVD. The numbers of analog and digital I/O modules connected to each ET200SP are different at ground potential and in the HVD, according to the specific needs. Each module is rated to withstand 707 V DC. The advantages of this choice are the following:
\begin{itemize}
    \item A better protection for the CPU, in the worst case that the max. 900 V DC in the capacitor banks of the cabinets at ground potential will be able to break through the insulation barriers of the electronic boards. 
    \item A better scalability of the system, in case further I/O modules are required.
    \item A reduction of the economical cost per channel, due to the only use of I/O modules for decentrated peripherals (i.e. ET200SP).
    \item A better maintainability of the system, thanks to the homogeneity of the system: the number of types of devices to be potentially replaced is now lower.
\end{itemize}
The ET200SP in the HVD, together with its IO modules and its SCALANCE XB004-1G, is visible on fig. \ref{fig:HVDreal}, top left.


\section{Conclusions}
A diagnostic neutral beam injector can represent a great opportunity to characterize the core of RFP plasmas. The basic redesign of the DNBI electrical systems, completed in 2024, was upgraded in 2025 for what concerns the most critical element, i.e. the High Voltage Deck.

Compared to the status of the previous HVD \cite{DNBI2025}, now the HVD is structured in a more rational, and safer way. The custom (triple) 20 kHz oil insulation transformer was substituted with a standard resin single phase one, at the expense of moving the capacitor bank for ARCPS inside the HVD.

The high voltage components for the 50 kV distribution feature TVS protection systems for a steeper response against overvoltages, more compact resistive dividers and a more complete set of shunts for V/I measurements and breakdown detection.

The detailed design of the ion source auxiliary power supplies, namely MIPS and GVPS, was completed. The electronic control boards were designed, tested and manufactured, ensuring more safety in terms of electric insulation of feedback signal and flexibility in communication with the control system. To improve the DNBI overall maintainability, the main control board and the "DriverIGBT" boards are intended to be multipurpose tools for several devices in the DNBI, namely MIPS, ARCPS and the voltage regulators and inverters serving the MHV. The boards data are shared with the DNBI community \cite{schematics}.

At last, the design of the entire DNBI PLC control system was upgraded to allow better scalability, maintainability and protection against overvoltages.

In parallel to the upgrade of the power supply systems, the vacuum pumping system is being upgraded with closed-cycle cryopumps, and the beam duct was redesigned and manufactured, promising lower reionization losses. In the next future, the DNBI ground potential cabinets hosting the inverters for the MHV will be upgraded for undelayable safety issues. The software for the control and data acquisition system will be developed. At last, the full commissioning of the DNBI is expected to be performed in 2027.

\section*{Acknowledgments}
This work has been carried out within the framework of Italian National Recovery and Resilience Plan (NRRP), funded by the European Union — NextGenerationEU (Mission 4, Component 2, Investment 3.1—Area ESFRI Energy—Call for tender No. 3264 of 28-12-2021 of Italian University and Research Ministry (MUR), Project ID IR0000007, MUR Concession Decree No. 243 of 04/08/2022, CUP B53C22003070006, ‘NEFERTARI - New Equipment for Fusion Experimental Research and Technological Advancements with Rfx Infrastructure’). Views and opinions expressed are however those of the author(s) only and do not necessarily reflect those of the European Union or the European Commission. Neither the European Union nor the European Commission can be held responsible for them.

The authors thank J. Stirling (Tokamak Energy Ltd), A. N. Karpushov and A. Listopad (EPFL) for the fruitful discussions and information exchange on DNBI technology.

\section*{Data availability}
The pdf schematics, bill of materials, KiCad and gerber files of the electronic boards that were described in the paper are available in ref. \cite{schematics}.

\bibliographystyle{myIEEEtran}
\bibliography{references}

\end{document}